\documentclass[letterpaper]{article} 
\usepackage{aaai2027}  
\usepackage[hyphens]{url}  
\usepackage{graphicx} 
\usepackage{natbib}  
\usepackage{caption} 
\usepackage{algorithm}
\usepackage{algorithmic}
\usepackage{amsmath}
\usepackage{amssymb}
\usepackage{amsthm}
\usepackage{placeins} 

\nocopyright

\usepackage{newfloat}
\usepackage{listings}
\DeclareCaptionStyle{ruled}{labelfont=normalfont,labelsep=colon,strut=off} 
\floatstyle{ruled}
\newfloat{listing}{tb}{lst}{}
\floatname{listing}{Listing}

\usepackage{booktabs}

\usepackage{multirow}

\title{SABRE: A Multi-Agent Approach for Selecting \\Out-of-Distribution Detectors Under a Budget}

\author{
    Mary Wisell,
    Salimeh Sekeh
}
\affiliations{
    San Diego State University\\
    mwisell4707@sdsu.edu, ssekeh@sdsu.edu
}

\begin{document}

\maketitle

\begin{abstract}
Post-hoc out-of-distribution (OOD) detection for vision--language models assumes that a detector chosen on a benchmark stays reliable once deployed. 
We show this fails across domains: on a single frozen encoder, a detector that leads in one domain can invert in another, scoring in-distribution inputs as more anomalous than genuine outliers, and the best detector changes from domain to domain, so no fixed choice is reliable throughout. 
We introduce SABRE (\underline{\bf S}elective \underline{\bf A}gentic \underline{\bf B}udgeted \underline{\bf R}eliability \underline{\bf E}nsemble), which replaces this fixed choice with per-regime selection at inference. 
Three language-model agents reason over a library of post-hoc detectors under a bounded query budget: a Selector chooses which detector to consult next, a Reporter consolidates the evidence for each input, and an Analyst calibrates detector reliability on a small labeled sample held out from the deployment domain and disjoint from the test data, weighting selection and aggregation without ever observing a scored input's label.
The library includes four multimodal density detectors we propose. 
Inferring the operating regime from data, SABRE tracks the strongest detector in each domain without prior knowledge of it, recovering reliable detection where a conventional detector inverts and converging to that detector where it is sound. 
A component analysis shows the agents are complementary: the Reporter's feedback yields consistent gains, and the Analyst's calibration is decisive against inversion, ruling out unreliable detectors so that aggregation no longer cancels the sound ones. 
Since no fixed rule can be trusted across domains, reliability must be established at deployment rather than assumed from a benchmark, and SABRE shows this can be done automatically.
\end{abstract}


\section{Introduction}
Deploying an out-of-distribution (OOD) detector for a vision--language model requires a choice: which post-hoc scoring rule to apply to the frozen encoder.
That choice is made once, guided by benchmark performance, and then fixed.
The detector that ranks highest on a standard benchmark is adopted and applied unchanged to whatever data deployment brings. 
Implicit in this practice is the assumption that a detector's ranking is a stable property of the encoder: that a rule which separates in- from out-of-distribution inputs well in one setting will continue to do so in another. 
There is little reason for this to hold. 
A post-hoc detector reads a particular structure of the encoder's representation, such as the spread of class scores, the geometry of the feature space, and the alignment between an image and a class name.
How well that structure separates novelty is a property of the domain as much as of the detector. 
When the deployment domain differs from the one a detector was selected on, the ranking that motivated the choice carries no guarantee.

This is not a hypothetical concern. 
On a single frozen encoder, a detector recommended by natural-image benchmarks can fail on a specialized domain not merely by underperforming but by inverting: scoring in-distribution inputs as more anomalous than genuine outliers, worse than random. 
The reversal is large, a detector near the top of one domain can sit near the bottom of another, and it is systematic: across several CLIP-family encoders and two markedly different domains, no single post-hoc detector is reliable throughout.
Deployment nonetheless forces one to be fixed in advance.

We recast OOD detection under domain shift as a selection problem: rather than commit to one rule, choose which detector to trust for the domain at hand, from evidence available at deployment. 
We realize this with SABRE, which coordinates a library of post-hoc detectors through three language-model agents under a bounded budget. 
The Selector (Llama~3.1~8B) chooses the next detector to run, or halts, from the evidence so far and a short description of each detector.
Its budget is counted in detector invocations, independent of hardware. 
The Reporter (Phi-3.5) consolidates the evidence for the current input, forming a fast per-input loop with the Selector. 
The Analyst (Phi-3.5) calibrates each detector's reliability on a small labeled sample held out from the deployment domain and disjoint from the test data, returning weights that steer selection and aggregation and corrective guidance for later inputs. 
No label of a scored input is ever seen.
This is the post-hoc calibration used to set a detector's threshold, raised to the level of choosing among detectors, and it assumes only a handful of labeled examples from the deployment stream, the minimum needed to identify a regime one cannot know in advance. 
The library pairs standard baselines with four multimodal density detectors we propose, each modeling where in-distribution data sits in the visual embedding space and adding a per-input caption signal, structured by semantic group, by a pooled covariance, by a cross-modal agreement penalty, and by a gated mixture.

Since the regime is inferred rather than assumed, SABRE tracks the strongest detector in each domain without being told which it faces. 
A component analysis activates the agents in turn: the Selector alone averages its choices with equal weight and sits near chance in an inverted regime, since averaging an inverted detector against a sound one cancels the signal.
The Reporter's feedback adds a small, consistent gain, and the Analyst's calibration is decisive, driving an inverted detector's weight toward zero so it is ruled out rather than corrupting the average, the largest improvement where fixed detectors fail, significant under a paired bootstrap and stable across validation and test, with corrective guidance refining the hardest regimes. 
SABRE thus recovers reliable detection in the domains where a fixed detector inverts, and matches the best available detector in the domains where one already suffices, without being told which domain it is in.

\paragraph{Contributions.}
\begin{itemize}\itemsep1pt
\item We establish that no fixed post-hoc OOD detector is reliable under domain shift: on an unchanged encoder, the leading detector inverts to below-chance where it was strongest.
\item We introduce SABRE, an agent-based framework that selects and reweights post-hoc detectors per regime at inference, under a bounded budget and a small held-out labeled sample.
\item We isolate each agent's role, showing reliability calibration is decisive against inversion while the Reporter and corrective guidance add consistent and margin-case gains.
\item We propose four multimodal density detectors combining visual-embedding geometry with a caption signal, broadening the pool the agents select from.
\end{itemize}

\section{Related Work}
Post-hoc OOD detection scores a frozen model without retraining. 
Early methods read the classifier response, such as maximum softmax probability \citep{hendrycks2017msp}, temperature-scaled input perturbation \citep{liang2018odin}, free energy \citep{liu2020energy}, and logit normalization \citep{wei2022logitnorm}, while others read the feature geometry through class-conditional Gaussians \citep{lee2018mahalanobis}, a relative-Mahalanobis correction \citep{ren2021rmd}, nearest neighbors \citep{sun2022knnood}, virtual-logit matching \citep{wang2022vim}, rectified activations \citep{sun2021react}, or sparsification \citep{sun2022dice}.
Deep ensembles \citep{lakshminarayanan2017ensembles} combine several models, and recent benchmarks and surveys standardize evaluation  \citep{yang2022openood,yang2024generalized}. 
Vision--language OOD detection adapts these ideas to contrastive encoders such as CLIP \citep{radford2021clip} and SigLIP \citep{zhai2023siglip}. 
Maximum concept matching scores an image by its similarity to class-name text prototypes \citep{ming2022mcm}, and extensions add pseudo-label captioners \citep{esmaeilpour2022zoc}, negation prompts \citep{wang2023clipn}, prompt learning \citep{miyai2023locoop}, mined negative labels \citep{jiang2024neglabel}, local matching \citep{miyai2024glmcm}, LLM-generated outlier classes \citep{cao2024eoe}, and analyses of fine-tuning \citep{ming2023finetune}. 
Each such method commits to one fixed rule whose reliability is reported on natural-image benchmarks. 
Our results show that this reliability does not transfer across domains, motivating selection over commitment. 
LLM agents extend language models with reasoning and acting \citep{yao2023react}, self-reflection \citep{shinn2023reflexion}, deliberate search \citep{yao2023tot}, and multi-agent collaboration \citep{wu2024autogen,hong2024metagpt}.
Language models have recently been surveyed as tools for anomaly and OOD detection \citep{xu2024llmood}, but to our knowledge have not been used to select detectors per regime under a budget. 
SABRE occupies this gap, coordinating post-hoc detectors with a team of language-model agents that calibrate reliability at deployment.

\section{Methodology}
\label{sec:methodology}
\paragraph{Problem formulation.}
Let $f$ be a frozen vision--language encoder mapping an image $x$ to a visual embedding $v=f(x)\in\mathbb{R}^{D}$, with $\{p_k\}_{k=1}^{K}$ the text embeddings of the $K$ in-distribution class names.
A post-hoc OOD detector is a scoring function $s\colon x\mapsto\mathbb{R}$ applied without modifying $f$, larger values indicating that $x$ is out-of-distribution, and detectors are compared by threshold-free AUROC.
Given a pool of $M$ detectors $\mathcal{S}=\{s_1,\dots,s_M\}$, standard practice fixes one $s_m$ on a benchmark and applies it unchanged, presuming that the ranking of detectors is stable across domains.
Writing $A_m(\mathcal{D})$ for the AUROC of $s_m$ on a domain $\mathcal{D}$, we instead seek a selection policy that, without knowing $\mathcal{D}$ in advance, attains per-domain performance close to $\max_m A_m(\mathcal{D})$ while bounding the worst case across domains.
The policy consults at most $B$ detectors per input (a query budget) and may use a small labeled calibration set $\mathcal{C}$ drawn from the deployment domain and disjoint from the evaluation data.
It never observes labels of the inputs it scores.

\paragraph{Detector pool.}
The pool combines established post-hoc baselines with four detectors we propose.
The baselines span the three standard mechanisms: confidence (MSP \citep{hendrycks2017msp}, Energy \citep{liu2020energy}), concept matching (MCM \citep{ming2022mcm}, and a static prompt-ensemble variant of T-QPM \citep{naiknaware2026tqpm} that we denote QPM), and feature-space density (Mahalanobis \citep{lee2018mahalanobis}).
The four proposed detectors share a form that adds feature-space density to a per-input caption signal,
\begin{equation}
s(x) \;=\; \log\!\bigl(1 + d(v)\bigr)\;+\;\gamma\,\bigl(1 - a(x)\bigr),
\label{eq:proposed-skeleton}
\end{equation}
where $d(v)$ is a density-based distance of $v$ from the in-distribution data, $a(x)\in[0,1]$ is the agreement between the image's caption and the nearest class name, and $\gamma\ge 0$ is a fixed weight balancing the caption term against the density term.
Both terms increase for stranger inputs, so neither can invert the score.
The members differ in how $d(v)$ is estimated: SMAP (Semantic Mahalanobis Anomaly Profile) uses a per-semantic-group covariance, RCAP (Regional Concentration Anomaly Profile) a single pooled covariance, MMCA (Multi-Modal Confidence Asymmetry) an explicit cross-modal penalty testing the caption-named group against the visually nearest group, and GMAP (Gated Mixture Anomaly Positioning) a gated mixture that modulates the caption term by the reliability of the text pathway.
They share the density baseline's covariance regularization, so comparisons isolate the choice of estimator.
Full formulations are in the supplementary materials.

\begin{figure*}[t]
\centering
\includegraphics[width=\textwidth]{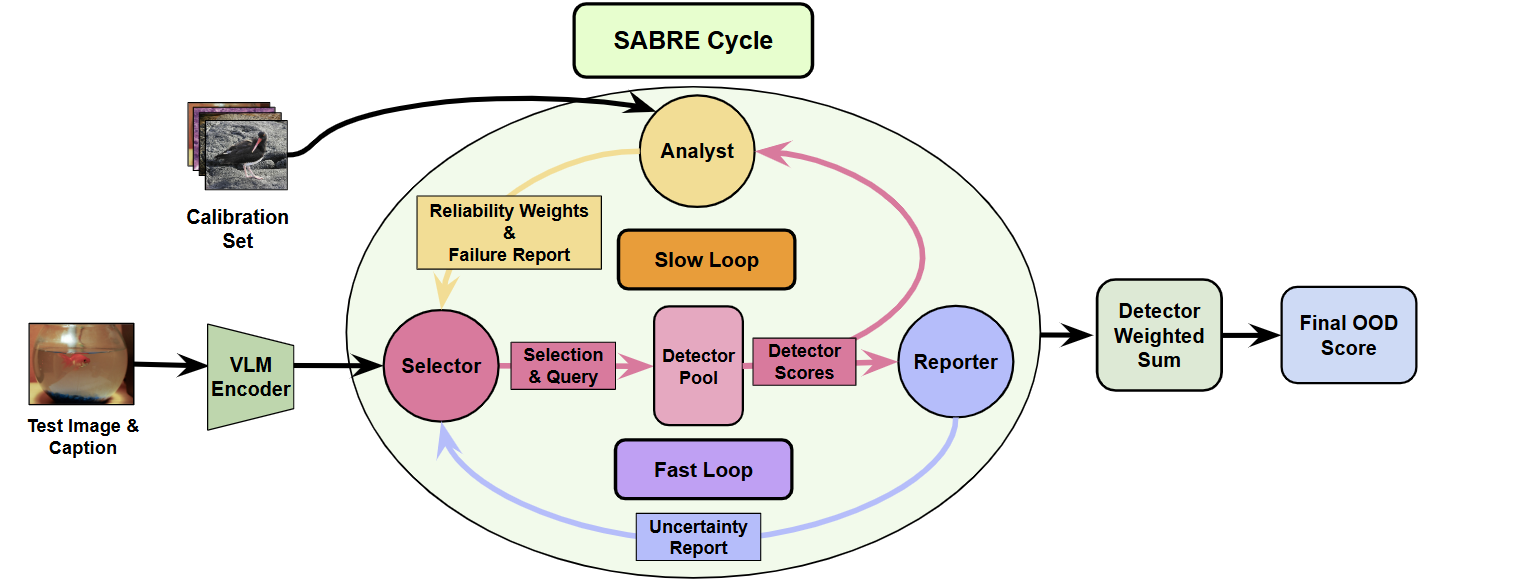}
\caption{SABRE. The Selector consults detectors from the pool one at a time under a budget $B$, the cycle repeating for the number of selections the budget permits, guided by the Reporter's per-image summary (fast loop, within one image) and the Analyst's reliability weights and failure taxonomy, estimated on the held-out calibration set (slow loop, across images). Scores are aggregated under the reliability weights to produce the final OOD score.}
\label{fig:sabre}
\end{figure*}

\subsection{SABRE}
SABRE makes the selection at inference through three language-model agents that coordinate the detector pool: a Selector that decides which detector to run next under the budget, a Reporter that consolidates the evidence gathered for the current image, and an Analyst that estimates detector reliability from the calibration set.
Figure~\ref{fig:sabre} shows the architecture and the information each agent exchanges.

\paragraph{Selector.}
For an input, the Selector chooses detectors to run one at a time.
At each step it receives the scores of detectors already run on this image, a one-line description of what each not-yet-run detector measures, the remaining budget, the Reporter's suggestion, and the Analyst's reliability weight, and it emits the next detector to run or a decision to stop, with a brief reason recorded in a shared trace.
Selection halts when the budget $B$ is exhausted or the evidence is judged sufficient, and the budget is counted in detector invocations rather than wall-clock time.

\paragraph{Reporter.}
After each detector call, the Reporter consolidates all evidence gathered for the current image into a short triage summary and a machine-readable note indicating which modality, visual or caption, remains uncertain.
This note returns to the Selector, forming a fast loop within a single image that refines the next choice.
The Reporter adds no detector calls, reorganizing evidence already obtained.

\paragraph{Analyst and the calibration mechanism.}
The Analyst operates a slow loop across images and supplies the mechanism that makes selection robust.
On the calibration set $\mathcal{C}$, held out from the deployment domain and never scored at test time, it measures each detector's reliability as its calibration AUROC $A_m(\mathcal{C})$ and converts this to a non-negative weight
\begin{equation}
w_m \;=\; \max\!\bigl(0,\; 2\,A_m(\mathcal{C}) - 1\bigr).
\label{eq:weight}
\end{equation}
A detector no better than chance on $\mathcal{C}$ ($A_m\le 0.5$) receives weight zero and is ruled out, so a detector that has inverted in the current domain is suppressed rather than allowed to corrupt the aggregate.
The final score is the reliability-weighted combination of the detectors run for the image,
\begin{equation}
s_{\mathrm{SABRE}}(x) \;=\;
\frac{\sum_{m\in\mathcal{R}(x)} w_m\,\tilde{s}_m(x)}
     {\sum_{m\in\mathcal{R}(x)} w_m},
\label{eq:aggregate}
\end{equation}
where $\mathcal{R}(x)$ is the set of detectors run for $x$ and $\tilde{s}_m$ is the score mapped to a common scale by its calibration statistics.
Because inverted detectors carry zero weight, aggregation no longer cancels sound detectors against unreliable ones, the failure mode of an unweighted average.
When a scored input's ground-truth label later becomes available the Analyst records the failure mode in a persistent taxonomy keyed by modality and caption content, and on subsequent images the Selector retrieves relevant entries as corrective guidance.
This guidance is the only component that adapts across images through experience rather than through the fixed calibration weights.

\paragraph{Why the components are complementary.}
The three agents address distinct failure points.
The Selector alone, aggregating with equal weight, is vulnerable in an inverted regime, where averaging a detector that points the wrong way against a sound one cancels the signal.
The Reporter improves which detectors are consulted but not how they are weighted.
The Analyst's calibration removes inverted detectors from the aggregate and is therefore decisive when a domain inverts a detector, while its corrective guidance refines the remaining ambiguous cases.
The Experiments section quantifies each contribution by activating the components in turn.

\begin{table*}[t]
\centering
\caption{Detector-level AUROC on CLIP ViT-B/16 (validation split, mean over five seeds). Best per dataset in \textbf{bold}. The confidence and concept-matching detectors (MSP, Energy, MCM, QPM) lead on natural images and invert below $50$ on histopathology, while the density detectors survive, and the leading detector changes between domains. No single row is safe across all columns. The other four encoders appear in the supplementary materials.}
\label{tab:detector}
\small
\begin{tabular}{lccccccccc}
\toprule
& \multicolumn{5}{c}{\textbf{Baselines}} & \multicolumn{4}{c}{\textbf{Ours}}\\
\cmidrule(lr){2-6}\cmidrule(lr){7-10}
Dataset & MSP & Energy & MCM & Maha. & QPM & SMAP & RCAP & MMCA & GMAP \\
\midrule
\multicolumn{10}{l}{\emph{Natural images (ImageNet-100 vs.)}}\\
iNaturalist & 87.9 & \textbf{98.5} & 97.7 & 93.1 & 88.3 & 92.9 & 93.0 & 91.9 & 92.5 \\
SUN         & 90.0 & \textbf{99.2} & 98.2 & 96.6 & 91.9 & 96.5 & 96.5 & 95.8 & 96.5 \\
Places      & 90.0 & \textbf{98.9} & 97.4 & 96.5 & 92.7 & 96.7 & 96.7 & 96.2 & 96.5 \\
Textures    & 94.6 & \textbf{99.0} & \textbf{99.0} & 95.4 & 84.5 & 93.9 & 94.0 & 93.0 & 95.4 \\
\midrule
\multicolumn{10}{l}{\emph{Histopathology}}\\
NCT-CRC     & 29.9 & 43.6 & 31.2 & \textbf{79.6} & 46.0 & 78.1 & 77.0 & 75.1 & 78.8 \\
CRC-VAL     & 34.9 & 67.7 & 37.2 & \textbf{70.8} & 49.2 & 68.0 & 68.9 & 65.5 & 70.5 \\
GCHTID      & 40.5 & 51.4 & 39.9 & 53.5 & 49.8 & \textbf{54.2} & 53.3 & 53.8 & 53.5 \\
\bottomrule
\end{tabular}
\end{table*}

\begin{table}[t]
\centering
\caption{Component analysis on histopathology (test split, $B=3$). Each column adds one mechanism. The reliability-calibration step (\textsc{+Mem.}) produces the recovery, and the best and worst single detector bound the range a fixed choice could occupy. $^{\ast}$ marks $p<0.001$ against \textsc{+Rep.}\ GCHTID is a difficulty ceiling.}
\label{tab:agent-histo}
\small
\setlength{\tabcolsep}{4pt}
\begin{tabular}{@{}llccccc@{}}
\toprule
Enc. & Dataset & Sel. & +Rep. & +Mem. & +Crit. & Best/Worst \\
\midrule
\multirow{3}{*}{B/16}
& NCT-CRC & 47.4 & 51.8 & \textbf{76.3}$^{\ast}$ & 76.2 & 79.6/32.9 \\
& CRC-VAL & 49.5 & 52.1 & \textbf{64.6}$^{\ast}$ & 64.4 & 69.6/40.1 \\
& GCHTID  & 43.6 & 45.7 & \textbf{51.4} & \textbf{51.4} & 52.1/40.7 \\
\midrule
\multirow{3}{*}{L/14}
& NCT-CRC & 68.5 & 71.1 & \textbf{84.4}$^{\ast}$ & 84.0 & 85.9/49.6 \\
& CRC-VAL & 73.0 & 75.3 & 81.5 & \textbf{81.9} & 80.4/50.5 \\
& GCHTID  & 55.1 & 55.5 & 56.2 & \textbf{58.0} & 60.7/49.6 \\
\bottomrule
\end{tabular}
\end{table}

\begin{table}[tb]
\centering
\caption{Selection controls on NCT-CRC (CLIP ViT-B/16, test). Reliability weighting lifts the pool by $10.1$ points over the unweighted combination, and random selection under the same weights matches SABRE, so the recovery comes from the calibration rather than the selection order. SABRE reaches $97\%$ of the weighted-pool quality at under a third of the calls. The best single detector is the strongest fixed detector for the domain, which a deployed system cannot identify in advance.}
\label{tab:controls}
\small
\begin{tabular}{lcc}
\toprule
Method & AUROC & Mean calls \\
\midrule
All detectors, unweighted       & 68.6 & 9.0 \\
All detectors, weighted         & 78.7 & 9.0 \\
Random selection (same weights) & 76.6 & 2.3 \\
\textbf{SABRE}                  & \textbf{76.3} & \textbf{2.6} \\
\midrule
Best single detector            & 79.9 & 1.0 \\
\bottomrule
\end{tabular}
\end{table}

\begin{table}[t]
\centering
\caption{The two costs, on NCT-CRC (CLIP ViT-B/16, test). Left: the query budget in detector calls per image, where quality plateaus at $B=3$ and the mean calls stay under the cap. Right: labeled calibration examples, where the four inverted detectors are ruled out across the range.}
\label{tab:cost}
\small
\begin{tabular}{cc@{\hskip 1.5em}ccc}
\toprule
\multicolumn{2}{c}{\emph{Budget}} & \multicolumn{3}{c}{\emph{Calibration size}}\\
\cmidrule(lr){1-2}\cmidrule(lr){3-5}
$B$ & AUROC & \# lab. & AUROC & ruled out \\
\midrule
1 & 73.4 & 50  & 78.1 & 4 \\
2 & 76.3 & 100 & 76.1 & 3 \\
3 & 76.3 & 150 & 76.3 & 4 \\
4 & 77.8 & 200 & 72.0 & 4 \\
5 & 77.4 & 250 & 75.7 & 4 \\
\bottomrule
\end{tabular}
\end{table}

\section{Experiments}
\label{sec:experiments}
Our evaluation proceeds in four steps. We first benchmark every detector on every dataset and encoder to establish that a fixed choice is unsafe.
We then activate SABRE's mechanisms one at a time to isolate which one recovers detection.
We check that this recovery generalizes across encoders, and we ask separately whether a domain-matched encoder removes the need for it.
Finally, two controls and two cost sweeps pin down what the recovery depends on and how little it needs.

\paragraph{Datasets.}
We use two domains that a single encoder represents very differently.
The natural-image domain takes ImageNet-100 \citep{deng2009imagenet,imagenet100kaggle} as in-distribution and iNaturalist \citep{vanhorn2018inaturalist}, SUN \citep{xiao2010sun}, Places \citep{zhou2018places}, and Textures \citep{cimpoi2014dtd} as out-of-distribution, following the standard large-semantic-space protocol \citep{huang2021mos}. 
The histopathology domain uses colorectal tissue: NCT-CRC \citep{kather2018nctcrc,nctcrckaggle} scores five in-distribution tissue types against held-out tissue classes, CRC-VAL \citep{kather2018nctcrc,kather2019crc,crcvalkaggle} repeats those classes on different scanners and sites as a covariate shift, and GCHTID \citep{lou2025gchtid,gchtidkaggle} is tissue from an unrelated source that none of the pool discriminates and that we report as a difficulty ceiling.
Each test scores $250$ in-distribution against $250$ out-of-distribution images on held-out splits that never overlap the calibration data.

\paragraph{Encoders and detectors.}
Five frozen CLIP-family encoders span general-purpose and specialized representations: CLIP ViT-B/16 (our primary encoder, and the one with the widest gap between the two domains) and ViT-L/14, SigLIP and SigLIP2 ViT-B/16, and the biomedical BiomedCLIP ViT-B/16, which we run on tissue only.
The pool of nine post-hoc detectors covers the three standard mechanisms: confidence (MSP, Energy), concept matching (MCM and QPM), and feature-space density (Mahalanobis and the four we propose, SMAP, RCAP, MMCA, GMAP).
We report AUROC as a percentage with $1000$-sample bootstrap intervals, and the detector benchmark averages five seeds.
SABRE runs with a budget of $B=3$ detector calls per image and calibrates on a labeled split ($30\%$ of the assembled samples, disjoint from the scored set).
Significance is assessed by a paired bootstrap on the shared samples.

\paragraph{Captions and preprocessing.}
The caption signal used by the four proposed detectors is produced once per image by a small open vision--language captioner and encoded by the same frozen text tower as the class prototypes, so no detector adds a trainable component.
All embeddings are extracted from frozen encoders under a fixed preprocessing pipeline, and a lightweight quality gate removes blank or saturated patches before scoring.
The calibration split is drawn from the deployment domain and is disjoint from every scored sample, so no number we report is computed on data the reliability weights were fitted on.
Full details are in the supplementary materials.

\subsection{Detector Benchmark}
\label{sec:exp-detector}
The detector benchmark, summarized for the primary encoder in Figure~\ref{fig:inversion} and Table~\ref{tab:detector} and for the remaining encoders in the supplementary materials, shows that the shift does more than reorder the detectors.
It inverts several of them.
MCM reaches $97.7$ on iNaturalist and falls to $31.2$ on NCT-CRC, Energy falls from $98.5$ to $43.6$, and the image-only MSP from $87.9$ to $29.9$.
A detector that inverts scores in-distribution tissue as more anomalous than genuine outliers, so its ranking is not merely weak but reversed.
The density detectors, Mahalanobis and the proposed SMAP, are the ones that hold up on tissue, and the strongest detector is different in each domain.
Since the same pattern recurs on all five encoders, no fixed choice is safe.
Whichever detector one commits to, there is a domain in which it fails, and a system that has to fix its choice before seeing the deployment data cannot tell which case it is in.

\begin{figure}[tbp]
\centering
\includegraphics[width=\columnwidth]{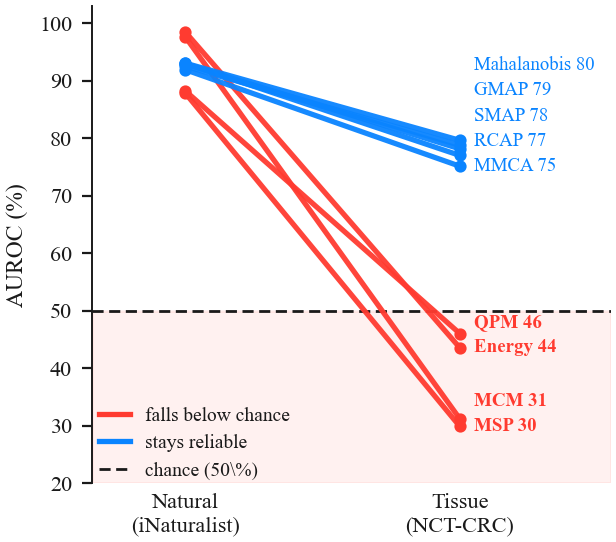}
\caption{Detector AUROC on CLIP ViT-B/16, from natural images (left) to tissue (right). Each line is one detector. The confidence and concept-matching detectors (red) drop from the top to below the chance line, scoring in-distribution tissue as more anomalous than genuine outliers, while the density detectors (blue) stay reliable. The best detector is not the same in the two domains.}
\label{fig:inversion}
\end{figure}


\subsection{Component Analysis}
\label{sec:exp-agent}
\begin{figure}[tbp]
\centering
\includegraphics[width=\columnwidth]{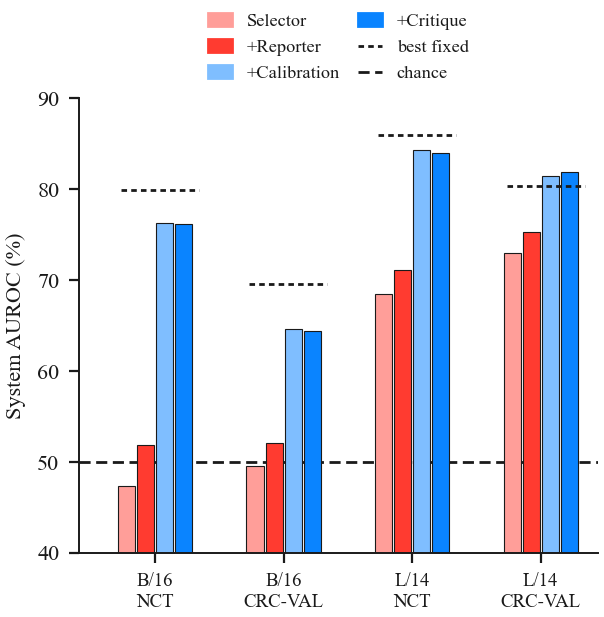}
\caption{Component analysis on the tissue cells where a fixed detector inverts (test split, $B=3$), as each mechanism is switched on. The two pre-calibration configurations (red) remain near chance. The reliability-calibration step (blue) recovers detection to near the best fixed detector (dashed), which the system is never told, and the recovery is largest exactly where a fixed detector fails.}
\label{fig:recovery}
\end{figure}

Activating SABRE's mechanisms in turn shows where the recovery comes from (Figure~\ref{fig:recovery}, Table~\ref{tab:agent-histo}). The Selector on its own, averaging its calls with equal weight, stays near chance in an inverted domain, because an inverted detector dragged into the average cancels a sound one.
Adding the Reporter changes little here.
The recovery arrives with the Analyst's reliability calibration, which drives the inverted detectors to zero weight and keeps the density detectors.
On NCT-CRC under CLIP ViT-B/16 this lifts the system from $51.8$ to $76.3$ (a gain of $24.5$ points, $p<0.001$), and under ViT-L/14 from $71.1$ to $84.4$ (a gain of $13.2$, $p<0.001$), reaching the neighborhood of the best fixed detector without ever being told which one it is.
Corrective guidance then sharpens the harder, covariate-shifted case, where on CRC-VAL under ViT-L/14 the full system reaches $81.9$, past the best fixed detector's $80.4$.

\begin{figure}[tbp]
\centering
\includegraphics[width=\columnwidth]{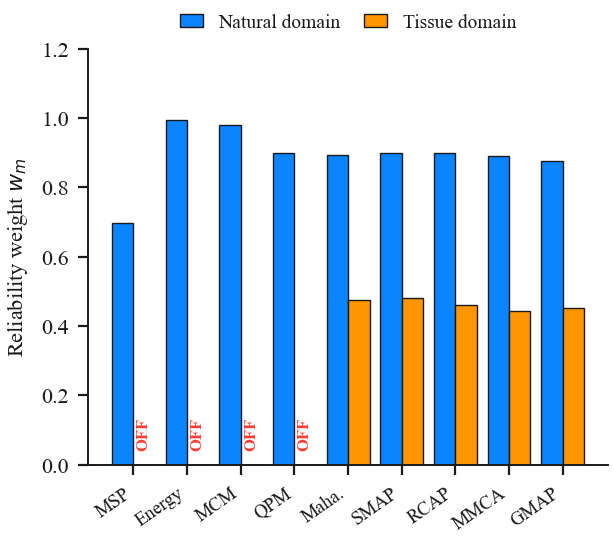}
\caption{The Analyst's reliability weights $w_m$ for the same system in the two domains (CLIP ViT-B/16). The confidence and concept detectors carry the weight on natural images and are switched off on tissue (marked \textsc{off}), and the density detectors reverse. The reweighting is inferred from a small labeled sample, not set by hand.}
\label{fig:reversal}
\end{figure}

The same configuration behaves differently where nothing is broken.
On natural images no detector inverts, and SABRE tracks the strongest fixed detector rather than trying to beat it, staying within about a point on every out-of-distribution set ($98.3$ against Energy's $97.7$ on iNaturalist, $98.9$ against MCM's $99.1$ on Textures).
Figure~\ref{fig:reversal} shows the reason directly. The Analyst's reliability weights flip between the two domains: the confidence and concept detectors that carry the signal on natural images are switched off on tissue, and the density detectors take over.
Inferring the domain from a small labeled sample and reweighting accordingly is exactly what a fixed rule cannot do.

\subsection{Generalization Across Encoders}
\label{sec:exp-backbones}
\begin{figure}[tbp]
\centering
\includegraphics[width=\columnwidth]{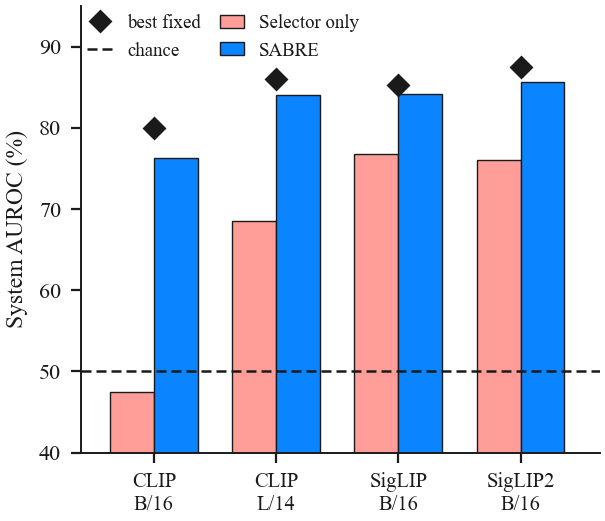}
\caption{Recovery across the general-purpose encoders (NCT-CRC, test). On every backbone the Selector alone (red) sits well below the best fixed detector (diamond), and full SABRE (blue) closes most of the gap.}
\label{fig:backbones}
\end{figure}

The recovery is not particular to one backbone.
Figure~\ref{fig:backbones} repeats the comparison on NCT-CRC for the general-purpose encoders.
In every case the Selector alone sits well below the best fixed detector and full SABRE closes almost the whole gap, from $47.4$ to $76.3$ on CLIP ViT-B/16 and from $76.1$ to $85.6$ on SigLIP2.
The gain scales with how far below the best detector the Selector starts, so the weakest starting point sees the largest recovery, and the effect holds regardless of which general-purpose encoder produces the features, ruling out an explanation specific to any one representation.
\subsection{A Domain-Specialized Encoder}
\label{sec:exp-biomed}
\begin{figure}[tbp]
\centering
\includegraphics[width=\columnwidth]{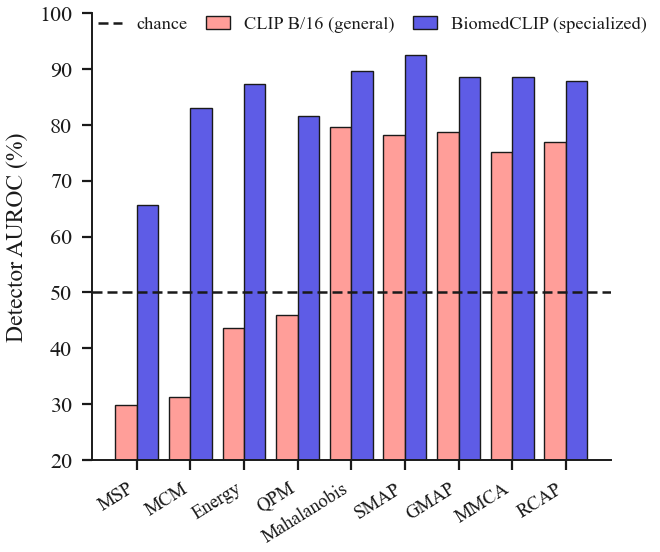}
\caption{Domain-specialized encoder ablation (NCT-CRC, test). BiomedCLIP (purple) lifts every detector over the general CLIP ViT-B/16 (red), but the confidence and concept detectors stay weakest and the density detectors stay strongest, so the ordering that motivates selection is unchanged.}
\label{fig:biomedclip}
\end{figure}

A natural question is whether the problem is an artifact of using a general-purpose encoder on tissue and would disappear under an encoder built for the domain.
We test this with BiomedCLIP, whose text encoder is trained on biomedical image--caption pairs (Figure~\ref{fig:biomedclip}).
The specialized encoder lifts every detector on tissue, and most of the weak ones, so the gap narrows, but the ordering does not change: the confidence and concept detectors (MSP, MCM, QPM) stay weakest and the density detectors strongest, and which density detector leads still varies by dataset.
The premise therefore survives a domain-matched encoder rather than being removed by it: a better encoder raises the floor for every detector but does not tell a practitioner which to trust, and SABRE still supplies that answer, lifting NCT-CRC from the Selector's $83.7$ to $92.5$, near the best fixed detector's $93.6$.
Selecting per regime and using a domain-matched encoder are complementary, not substitutes.

\subsection{What the Recovery Depends On}
\label{sec:exp-controls}
Two controls separate measuring reliability from selecting under a budget (Table~\ref{tab:controls}).
Combining the whole pool unweighted is worse than useless, reaching only $68.6$, below the best single detector, because the inverted members cancel the sound ones. Weighting the same nine detectors by their measured reliability lifts this to $78.7$, a gain of $10.1$ points from calibration alone.
Replacing the learned Selector with uniform-random choice under those weights yields $76.6\pm0.2$, level with SABRE's $76.3$, so the recovery comes from measuring reliability at deployment, not from the order in which the correlated sound detectors are consulted.
What selection adds is cost: SABRE reaches $97\%$ of full-pool quality while calling $2.6$ detectors per image instead of nine, and routes to the right detectors when the domain does not need rescuing.
Both costs are small (Table~\ref{tab:cost}). Quality plateaus by a budget of three calls, with the mean spent staying below the cap, and the four inverted detectors are ruled out across the full range of calibration sizes, down to fifty labeled examples, so a new domain needs only a small labeled sample rather than a full validation set.


\section{Discussion}
\label{sec:discussion}
These results reframe post-hoc OOD detection under domain shift as a selection problem rather than a design problem.
What governs reliability at deployment is not which detector scored best on a benchmark but which remains reliable on the domain at hand, and that is measurable from a small labeled sample without touching the test data: once measured, an inverted detector is assigned zero weight and the aggregate stops canceling sound detectors against unsound ones, so calibration at deployment, rather than any particular detector, carries the recovery.
A practitioner facing a new domain therefore need not guess which post-hoc rule will transfer.
SABRE turns that guess into a measurement, spends a bounded number of detector calls per input, and recovers reliable detection automatically, converging to the strongest available detector where one already works, rescuing detection where the benchmark-preferred detector inverts, and drawing on the four multimodal density detectors we introduce to stay reliable where confidence and concept-matching rules fail.

Two properties set the method's scope.
SABRE tracks the best single detector rather than surpassing it, so its value lies in reaching that level without prior knowledge of which detector attains it, and where no detector in the pool separates the classes, as on the hardest tissue source, selection cannot create signal that is absent and the pool itself is the object to strengthen.
These boundaries point to natural extensions: larger and more diverse detector pools, calibration that adapts online as labeled data accrues, and selection over modalities beyond image and caption.

\section{Conclusion}
\label{sec:conclusion}
Post-hoc OOD detectors for vision--language models do not carry their benchmark reliability across domains: on a frozen encoder the detector that leads on natural images can fall below chance on a specialized domain, and the strongest detector changes from one domain to the next.
SABRE meets this by selecting and reweighting detectors at deployment, coordinating a pool through three language-model agents that calibrate reliability on a small labeled sample and consult detectors under a bounded budget.
It recovers reliable detection where a fixed detector inverts, matches the best available detector where one already suffices, and does so with a handful of labeled examples and a few detector calls per input.
Reliability under domain shift is therefore not a property to assume from a benchmark but one to establish at deployment, and doing so can be automated.

\paragraph{Use of Generative AI}
The authors used generative AI tools (large language models) to assist with copyediting, spelling and grammar correction, improving clarity, condensing text to meet the length limit, and checking for wording or content that could compromise double-blind anonymity. All research contributions, including the study design, experiments, analyses, and figures, were produced and verified by the authors, who take full responsibility for the entire content of the paper, including all text, figures, and references.

\paragraph{Acknowledgments}
This work has been partially supported by NSF CAREER CCF-2451457. The findings are those of the authors only and do not represent any position of these funding bodies.

\bibliography{arxiv_references}
\clearpage
\appendix
\setcounter{secnumdepth}{2}
\begin{center}\Large\bfseries Supplementary Material\end{center}
\section{Data Preprocessing and Reproducibility}
\label{sec:supp-preprocessing}
This section documents the full preprocessing pipeline so that the results can be reproduced. Every vision--language encoder is used frozen. The only trained component is a linear probe head fit on cached embeddings (Section~\ref{sec:supp-linprobe}), and the encoders are never unfrozen at any stage.
\subsection{Encoders and Embedding Extraction}
The five backbones are loaded through the \texttt{open\_clip} library and held frozen, with all parameter gradients disabled and the models in evaluation mode: CLIP ViT-B/16 and ViT-L/14 (OpenAI weights, $512$- and $768$-dimensional), SigLIP and SigLIP2 ViT-B/16 (WebLI weights, $768$-d), and the domain-specialized BiomedCLIP ViT-B/16 ($512$-d). BiomedCLIP is loaded through the pretrained-model constructor rather than the standard one, because it bundles its own preprocessing configuration inside the checkpoint. SigLIP2 differs from the other general-purpose encoders in that its pretraining adds a captioning objective alongside the contrastive loss, aligning its text and image spaces more tightly, which is why we include it when probing whether the caption channel helps.
Each image is passed through the encoder's own preprocessing transform, which resizes with bicubic interpolation and center-crops to the native input side ($224$ px for all five backbones) and normalizes with that encoder's channel statistics. The visual encoder then produces an embedding that is $L_2$-normalized. A lightweight quality gate rejects inputs below $64$ px per side, or with a mean intensity outside $[10, 240]$ or a standard deviation below $5$ (on the $0$ to $255$ scale), which removes blank and saturated patches. All embeddings are precomputed once per encoder and cached, so every detector and every agent run reads identical features.
\subsection{Class Prototypes and Prompt Ensembles}
Class-name prototypes are built by prompt ensembling: for each class, every template is filled with the class description, encoded by the frozen text encoder, and the resulting embeddings are averaged and $L_2$-normalized. Natural-image classes use five generic templates (``a photo of a \{\}'', ``a photograph of a \{\}'', ``an image of a \{\}'', ``a picture of a \{\}'', ``a photo of the \{\}''), and tissue classes use five domain templates (``a histopathology image of \{\}'', ``a microscopy image showing \{\}'', ``an H\&E stained tissue patch of \{\}'', ``a pathology slide showing \{\}'', ``a high-magnification view of \{\}'').
The tissue prototypes encode the descriptive forms of the abbreviations rather than the abbreviations themselves: ADI is ``adipose tissue'', MUS ``smooth muscle'', STR ``cancer-associated stroma'', and NORM and TUM take dataset-appropriate forms (``normal colorectal mucosa'' and ``colorectal tumour epithelium'' for the colorectal data, ``normal gastric mucosa'' and ``gastric tumour epithelium'' for the gastric data). The QPM baseline scores against four template-specific prototype banks, each built from a distinct template through the same text tower, rather than against a single averaged bank.
\subsection{Captions and the Caption Signal}
Per-image captions are generated once, offline, by a small open vision--language captioner MiniCPM-V from Ollama, on the $224 \times 224$ H\&E patches. 
TIFF images are converted to PNG in memory before captioning, because the captioner does not accept TIFF input, and the file on disk is never modified. 
Captioning is resumable, keyed on the (image, dataset, split) triple. For each patch the captioner returns a short morphological description together with a one-sentence texture-and-color summary, and the two are concatenated into a single string before text encoding. 
Three guards keep the caption from leaking the label. 
The captioning prompt explicitly forbids naming or guessing the tissue type, organ, or diagnosis, and asks only for visual morphology, staining, and texture. 
Captioners with medical or pathology pretraining are disallowed as the caption source, since such a model could name the class from priors rather than from pixels.
Concretely, the pipeline refuses pathology- or biomedical-pretrained captioners, such as LLaVA-Med, BiomedCLIP, PathCLIP, CONCH,
UNI, BioCLIP, at runtime and admits only general-purpose open captioners such as MiniCPM-V, LLaVA, or Moondream2. 
Every class label used anywhere in the pipeline comes from the directory structure, never from the captioner. 
The caption is used only by the Reporter and by the caption term of the four proposed detectors. 
It never sets a detector's threshold or labels a scored image. 
Natural images use a scene-content prompt (main subject, colors, spatial layout) in place of the histology prompt. 
Each caption is encoded by the same frozen text tower as the class prototypes and L2-normalized, giving a caption embedding $c$ from which the caption-side class posterior $p_t = \mathrm{softmax}(Pc/\tau)$ is formed. 
When a patch has no caption, the caption term is dropped and the density term is used alone.
\subsection{Confidence Baselines: Linear Probe}
\label{sec:supp-linprobe}
The confidence baselines (MSP, Energy) read the logits of a single linear layer trained on the frozen in-distribution training embeddings of the known classes, following the standard CLIP linear-probe protocol (Adam, learning rate $10^{-3}$, up to $50$ epochs, with the best-validation checkpoint retained). This head is the only trained component in the pipeline. It is fit entirely from the cached embeddings, the encoder is never unfrozen, and the held-out out-of-distribution tissue classes are excluded from its training.
\subsection{Datasets, Splits, and Scoring Protocol}
The natural-image in-distribution set is ImageNet-100 ($100$ fixed synset IDs), with iNaturalist, SUN, Places, and Textures as out-of-distribution sets. The histopathology data is NCT-CRC-HE-100K (the primary tissue benchmark), its companion CRC-VAL-HE-7K from a disjoint cohort scanned and stained differently (a covariate shift), and GCHTID, an independent gastric histopathology source that none of the pool discriminates and that we report as a difficulty ceiling. The tissue patches carry eight classes, of which five (ADI, MUS, NORM, STR, TUM) are treated as in-distribution and three (DEB, LYM, MUC) as held-out out-of-distribution, and a background class is skipped entirely.
Splits are created with a fixed seed under a $70/15/15$ train/validation/test partition. Every class is capped to the size of the smallest class so all classes contribute equally, and counts are rounded down to multiples of ten at each split boundary. The three out-of-distribution tissue classes are split and stored identically to the in-distribution ones but are excluded from linear-probe training and from prototype building. NCT-CRC and GCHTID each build their own tissue prototypes with dataset-appropriate descriptions, while CRC-VAL, which shares the taxonomy, is scored against the NCT prototypes so that only the scanner and stain differ. Embeddings are stored per split.
For every evaluation we assemble a scored set of $500$ images, $250$ in-distribution against $250$ out-of-distribution, drawn from the held-out test split. The Analyst's calibration set is a further disjoint labeled sample drawn from the same deployment domain: at the default $30\%$ fraction it is $150$ images ($75$ per class-balanced half), and it never overlaps the scored set. The detector benchmark averages five seeds ($1000$ to $1004$), bootstrap intervals use $1000$ resamples, and paired significance uses $2000$. Splitting and sampling are controlled by fixed seeds so that a rerun reproduces the same partitions.

\subsection{Computing Infrastructure}
All experiments were completed with an NVIDIA GeForce RTX~5080 Laptop GPU (16~GB), using Python~3.12.0.

\section{Extended Related Work}
\label{sec:extended-related}
\subsection{Post-hoc Out-of-Distribution Detection}
Post-hoc detectors attach an OOD score to a frozen, already-trained model, avoiding the cost and instability of retraining.
The earliest and still most-cited reads the classifier's response: the maximum softmax probability \citep{hendrycks2017msp} treats low predicted confidence as evidence of novelty, ODIN \citep{liang2018odin} sharpens it with temperature scaling and input perturbation, the energy score \citep{liu2020energy} replaces the maximum probability with the log-sum-exp of the logits, and logit normalization \citep{wei2022logitnorm} addresses the overconfidence that undermines such scores.
A second family reads the geometry of the feature space rather than the head: the Mahalanobis detector \citep{lee2018mahalanobis} fits class-conditional Gaussians with a tied covariance and scores by distance to the nearest class mean.
A relative-Mahalanobis correction \citep{ren2021rmd} improves near-OOD behavior.
Deep nearest-neighbor distance \citep{sun2022knnood} is a non-parametric alternative, and virtual-logit matching \citep{wang2022vim} combines a feature-space residual with the class logits.
Orthogonal refinements act on the representation before scoring, truncating extreme activations \citep{sun2021react} or sparsifying weights \citep{sun2022dice}.
Deep ensembles \citep{lakshminarayanan2017ensembles} aggregate several independently trained models.
Standardized benchmarks and surveys \citep{yang2022openood, yang2024generalized} organize these into confidence-, energy-, and distance-based groups and expose their sensitivity to the evaluation setting.
A recurring assumption is that a detector's effectiveness, established on a benchmark, is a stable property one can rely on at deployment.
Our work questions that assumption directly, and the density-based members of this family are the ones that survive the domain shift we study.
\subsection{OOD Detection for Vision--Language Models}
Contrastive image--text encoders such as CLIP \citep{radford2021clip} and SigLIP \citep{zhai2023siglip} enable zero-shot recognition by comparing an image embedding to text embeddings of candidate class names, which immediately suggests a zero-shot OOD score.
Maximum concept matching (MCM) \citep{ming2022mcm} formalizes it: the text embeddings of the in-distribution class names act as concept prototypes, and an image far from every prototype is flagged as OOD.
Subsequent methods enrich the text side of the comparison.
ZOC \citep{esmaeilpour2022zoc} trains a captioner to propose candidate OOD labels.
CLIPN \citep{wang2023clipn} learns explicit ``no'' prompts so the encoder can express negation.
LoCoOp \citep{miyai2023locoop} uses prompt learning with local regularization in a few-shot regime.
NegLabel \citep{jiang2024neglabel} mines a large set of negative labels from a corpus.
GL-MCM \citep{miyai2024glmcm} adds local matching for images with multiple objects, and EOE \citep{cao2024eoe} uses a large language model to envision outlier classes.
A study of fine-tuning \citep{ming2023finetune} characterizes how adapting the encoder affects OOD behavior.
What these approaches share is a dependence on the encoder's text representation of the label space and a single fixed rule selected on natural-image benchmarks.
Our experiments show that when the deployment domain is one the encoder represents poorly, methods of this kind can not only weaken but invert, and that the identity of the best detector is domain-dependent, so the fixed-rule assumption inherited from single-modal detection is the precise point of failure.
\subsection{Language-Model Agents}
A parallel line of work turns language models into agents that interleave reasoning with action.
ReAct \citep{yao2023react} alternates chain-of-thought with tool or environment calls.
Reflexion \citep{shinn2023reflexion} adds verbal self-critique stored and reused across attempts.
Tree-of-Thoughts \citep{yao2023tot} searches over intermediate reasoning states.
Multi-agent frameworks coordinate several such agents: AutoGen \citep{wu2024autogen} structures them as conversable roles exchanging messages, and MetaGPT \citep{hong2024metagpt} assigns standard-operating-procedure roles to a collaborative pipeline.
These systems are typically evaluated on open-ended reasoning, coding, or tool-use tasks.
Their relevance here is mechanistic rather than domain-specific: SABRE's roles, a Selector that acts under a budget, a Reporter that summarizes intermediate evidence, and an Analyst that reflects on held-out outcomes to calibrate future behavior, mirror the acting, reporting, and reflecting patterns these works establish, but are directed at a measurement problem rather than a generation one.
\subsection{Language Models for OOD and Detector Combination}
A recent survey catalogs the growing use of language models for anomaly and OOD detection \citep{xu2024llmood}, largely as scorers or explainers of individual inputs, and some CLIP-based methods already invoke a language model to generate auxiliary labels \citep{cao2024eoe}.
This is distinct from using agents to choose among detectors.
Classical detector combination, such as ensembling or averaging several scores, as with deep ensembles \citep{lakshminarayanan2017ensembles}, treats the pool as fixed and queries all of it, whereas SABRE selects a subset under a call budget and weights it by measured reliability, declining to spend budget on detectors that calibration shows to be unreliable in the current regime.
To our knowledge, per-regime, budget-constrained selection among post-hoc detectors, driven by held-out reliability estimates rather than a fixed combination rule, has not been studied for vision--language OOD detection.
\subsection{Benchmarks and Encoders}
We evaluate across two markedly different domains. For natural images we follow the standard large-semantic-space protocol \citep{huang2021mos}, with out-of-distribution sets drawn from iNaturalist \citep{vanhorn2018inaturalist}, the SUN database \citep{xiao2010sun}, Places \citep{zhou2018places}, and Describable Textures \citep{cimpoi2014dtd}.
For the specialized domain we use colorectal histopathology: NCT-CRC-HE-100K \citep{kather2018nctcrc} provides 100,000 H\&E-stained patches across nine tissue classes, and the associated CRC-VAL-HE-7K set, from a disjoint patient cohort, supports evaluation under covariate shift, and both derive from the study of \citet{kather2019crc}.
To test whether our findings depend on the encoder, we evaluate several CLIP-family backbones alongside the domain-specialized BiomedCLIP \citep{zhang2023biomedclip}, whose text encoder is trained on biomedical image--caption pairs.
The contrast between a general-purpose encoder on tissue and a specialized one is what isolates the non-transfer of a fixed detector's reliability.

\section{Detector Formulations}
\label{sec:supp-detectors}
This section gives the full formulation of the four proposed detectors and the five baselines, as implemented. The baseline QPM is a static prompt-ensemble variant of T-QPM \citep{naiknaware2026tqpm} that uses fixed class-prototype prompts in place of the original's query-time prompting, and we write QPM for it throughout. Every detector operates on a frozen encoder that produces an $L_2$-normalized visual embedding $v\in\mathbb{R}^{D}$, a matrix of $K$ $L_2$-normalized class-name prototypes $P\in\mathbb{R}^{K\times D}$, and, when a caption is available, an $L_2$-normalized caption embedding $c\in\mathbb{R}^{D}$ obtained by encoding a per-image caption with a small open vision--language captioner. Because all vectors are unit norm, the products $Pv\in\mathbb{R}^{K}$ and $Pc\in\mathbb{R}^{K}$ are vectors of cosine similarities to the class prototypes.

\subsection{Shared Estimators}
\paragraph{Shrinkage covariance.}
Every Mahalanobis term uses a Ledoit--Wolf-style shrinkage estimator for each sample covariance $S$ formed from $n$ fitting points,
\begin{equation}
\Sigma = (1-\alpha)\,S + \alpha\,\frac{\operatorname{tr}(S)}{D}\,I,
\qquad
\alpha = \operatorname{clip}\!\Bigl(\tfrac{\lVert S - \bar\mu I\rVert_F^{2}}{n\,\lVert S\rVert_F^{2}},\,0,\,1\Bigr),
\label{eq:supp-shrink}
\end{equation}
with $\bar\mu=\operatorname{tr}(S)/D$, followed by inversion of $\Sigma + 10^{-6} I$. The Mahalanobis baseline and the proposed density detectors share this estimator and the same inversion, so any difference among them is attributable to how the fitting points are partitioned rather than to covariance regularization.

\paragraph{Semantic groups.}
Three of the proposed detectors partition the $K$ class prototypes into $G\le K$ semantic groups by agglomerative merging on prototype cosine similarity: starting from singletons, the two clusters with the highest mean pairwise similarity are merged until $G$ groups remain (default $G=4$). The procedure is deterministic and adds no external clustering dependency. Each training visual embedding is then assigned to the group of its nearest prototype, $\arg\max_k (Pv)_k$.

\paragraph{Caption agreement.}
When a caption is present, its agreement with the known classes is the largest cosine similarity between the caption embedding and any class prototype,
\begin{equation}
a(x) = \max_k (Pc)_k .
\label{eq:supp-caption-agree}
\end{equation}
This is a similarity rather than a softmax posterior, and it is small when the caption resembles no known class name. The captioner is prompted so as not to name or guess the class, so $a(x)$ reflects a generic visual description rather than a class assertion.

\subsection{Profile Detectors: SMAP, RCAP, MMCA}
The three profile detectors share a common form. Writing $d(v)$ for a visual density distance,
\begin{equation}
s(x) = \underbrace{\log\!\bigl(1+d(v)\bigr)}_{\text{density}} \;+\; \gamma\underbrace{\bigl(1-a(x)\bigr)}_{\text{caption deficit}},
\qquad \gamma = 2,
\label{eq:supp-skeleton}
\end{equation}
with the caption deficit omitted, and the density term used alone, whenever no caption is available. Both terms are non-negative and grow for more novel inputs, so the composite cannot invert through cancellation between them, and with no caption the score is a pure density readout rather than a concept-match score.

\paragraph{SMAP (Semantic Mahalanobis Anomaly Profile).}
A separate mean $\mu_g$ and shrinkage covariance $\Sigma_g$ are fit per semantic group over the training embeddings assigned to it, and groups with fewer than two points are dropped. The density is the minimum Mahalanobis distance over groups,
\begin{equation}
d(v) = \min_g\,(v-\mu_g)^{\top}\Sigma_g^{-1}(v-\mu_g),
\label{eq:supp-smap}
\end{equation}
so each region of the space keeps its own shape. The profile handed to the Selector records the per-group distances, the nearest group and its member classes, the caption match, and a signature naming which channel (visual, caption, or both) carries the anomaly.

\paragraph{RCAP (Regional Concentration Anomaly Profile).}
RCAP mirrors SMAP but pools a single covariance $\Sigma$ across all groups. The per-group residuals are concatenated and one shrinkage covariance is fit, giving
\begin{equation}
d(v) = \min_g\,(v-\mu_g)^{\top}\Sigma^{-1}(v-\mu_g).
\label{eq:supp-rcap}
\end{equation}
Sharing one covariance borrows statistical strength across groups and is better conditioned when a group has few samples, so it is a distinct estimator rather than a reparameterization of SMAP. RCAP additionally reports a scale-free decisiveness signal in its profile, the relative separation $(d_{(2)}-d_{(1)})/(d_{(1)}+\epsilon)$ between the nearest and second-nearest group distances together with the caption's top-two margin. These enter the Selector's evidence but not the score of Eq.~\eqref{eq:supp-rcap}.

\paragraph{MMCA (Multi-Modal Confidence Asymmetry).}
MMCA uses the same per-group covariances as SMAP for its density and adds an explicit cross-modal coupling penalty. Let $g_t$ be the group of the class whose prototype the caption matches best, that is the group containing $\arg\max_k (Pc)_k$, and let $d_{g_t}$ be the sample's Mahalanobis distance to that group. The penalty is the extra log-distance from the caption-named group beyond the nearest group,
\begin{equation}
\kappa(x) = \max\!\Bigl(0,\ \log\!\bigl(1+d_{g_t}\bigr) - \log\!\bigl(1+\textstyle\min_g d_g\bigr)\Bigr),
\label{eq:supp-mmca-couple}
\end{equation}
and the score is
\begin{equation}
\begin{split}
s(x) ={}& \log\!\bigl(1+\textstyle\min_g d_g\bigr) + \delta\,\kappa(x) + \gamma\bigl(1-a(x)\bigr),\\
        & \delta = 0.25,\ \gamma = 2.
\end{split}
\label{eq:supp-mmca}
\end{equation}
The penalty is zero when the caption points at the group the image is already nearest and grows when the caption names a group the image is far from. It carries a deliberately small weight $\delta$ because the caption's group hypothesis is trustworthy only where the text encoder is grounded in the domain. This coupling is the only genuinely cross-modal quantity among the four detectors, since neither channel produces it alone.

\subsection{GMAP: Gated Mixture with a Learned Text Gate}
\label{sec:supp-gmap}
GMAP belongs to the same family as the profile detectors, a feature-space density combined with a per-image caption signal, both terms increasing for more novel inputs so that neither can invert the score, but instantiates the two terms differently from Eq.~\eqref{eq:supp-skeleton}: a per-class mixture density in place of the per-group Gaussian, and a learned gate over a caption-space density in place of the fixed additive caption deficit. This is the sense in which, as the main paper puts it, GMAP is the family member that modulates the caption term by the reliability of the text pathway.

\subsection{Baseline Detectors}
The pool also contains five post-hoc baselines on the same frozen embeddings, each returning a score for which larger is more out-of-distribution.

\paragraph{MSP.}
Maximum softmax probability on the visual embedding \citep{hendrycks2017msp}, $s = 1-\max_k \mathrm{softmax}(Pv/\tau)_k$, with $\tau$ the encoder temperature. Image only.

\paragraph{Energy.}
The free energy of the class logits \citep{liu2020energy}, with the leading temperature factor retained so the magnitude matches the published definition,
\begin{equation}
s = -\,\tau\,\log\!\sum_k \exp\!\bigl((Pv)_k/\tau\bigr).
\label{eq:supp-energy}
\end{equation}
Dropping the factor is a positive rescale that leaves AUROC unchanged but makes the value no longer the energy the reference defines. Image only.

\paragraph{MCM.}
Maximum concept matching \citep{ming2022mcm}, $s = 1-\max_k \mathrm{softmax}(Pv/T)_k$. The softmax over cosine similarities makes the score depend on how similarity mass is spread across all classes, not only on the top similarity. The temperature $T$ is a method hyperparameter distinct from the encoder temperature, with default $T=1$, and may be tuned on a validation split. Image only.

\paragraph{Mahalanobis.}
One Gaussian per known class with a single tied covariance \citep{lee2018mahalanobis}, scored by the nearest-class distance $s = \min_k (v-\mu_k)^{\top}\Sigma^{-1}(v-\mu_k)$. Class assignment uses ground-truth training labels where available, with the same nearest-prototype fallback and the same shrinkage estimator (Eq.~\ref{eq:supp-shrink}) as the proposed density detectors, so the only remaining difference from RCAP is per-class rather than per-group Gaussians. We use the standard simplified form and omit the input preprocessing and multi-layer feature ensemble of the original, neither of which transfers to a frozen encoder scored on cached final-layer embeddings. Image only.

\paragraph{QPM.}
A static prompt-ensemble matcher \citep{naiknaware2026tqpm}. Four class-prototype banks are built from four distinct prompt templates. Two banks are matched against the visual embedding and two against the caption embedding, each contributing its maximum cosine similarity $Q_i=\max_k \bigl(\mathrm{proto}^{(i)}\!\cdot e\bigr)_k$ for the appropriate embedding $e$. The score is
\begin{equation}
s = 1 - \sum_{i=1}^{4}\beta_i\,Q_i,
\label{eq:supp-qpm}
\end{equation}
with weights $\beta_i$ that default to a uniform $0.25$ and may be tuned on validation. The four banks must come from four different templates, since identical banks collapse the weights and reduce the method to an equal mixture of concept matching and the caption matcher. The implementation guards against this at construction. This static re-implementation drops the temporal-consistency and covariance-update terms of the original T-QPM so that it scores a single sample and is directly comparable to MCM.

\subsection{Relationship Among the Detectors}
The four proposed detectors form one family that varies along three axes: how the covariance is partitioned (per group, pooled, or per-class mixture), whether an explicit cross-modal penalty is present, and single-component versus mixture density. Their scores are strongly rank-correlated, with Kendall $\tau$ between $0.72$ and $0.99$ across settings, so we treat them as variants of a single approach rather than as independent detectors. Their value in the pool is that their reliability varies by domain, which gives the Selector a spectrum to choose from.

\section{Agent Implementation Details}
\label{sec:supp-agents}
This section describes the three agents, the shared interface through which they see the detector pool, and the pipeline that runs them, as implemented.

\subsection{Models and Decoding}
The Selector runs on Llama~3.1~8B, and the Reporter and Analyst on Phi-3.5, all served locally from cached weights. All agents decode greedily at temperature $0$, with top-$k=1$, top-$p=1$, and a fixed seed set explicitly, so a rerun returns the same text rather than merely a similar one and a run is reproducible up to hardware-level floating-point nondeterminism. Generation is capped per role (Table~\ref{tab:supp-agent-config}): the Selector emits a short structured decision and is capped at $96$ tokens, while the Reporter's summary is read by the Selector on the next step and is held at $256$ tokens so it is not truncated mid-sentence, which would corrupt that input. When an agent is asked for JSON, generation additionally stops at the first blank line or code fence, so tokens are not spent on trailing prose that the parser discards. A malformed or failed agent call falls back rather than halting the run: the Selector defaults to a fixed priority order over the not-yet-run detectors, and a failed Analyst call leaves the taxonomy unchanged for that image.

\subsection{Shared Detector View}
The Selector and Reporter never read a detector's raw profile directly. Each detector's method-specific profile is passed through a shared adapter that maps it to a common view: the method name, its calibrated score, its own failure signature, an inferred dominant modality (visual, text, cross-modal, or neither), and a short list of human-readable evidence strings drawn from whatever structured fields the detector exposes. This gives both agents a single set of fields to reason over regardless of which detector produced the evidence. The five baselines, which emit only a scalar and a small diagnostic dictionary, are wrapped by a second adapter that presents the same scoring interface as the proposed detectors and returns a deliberately thin profile marked as scalar-only. A baseline therefore competes for selection on the same footing as a proposed detector, and the Selector can see that it offers no decomposed structure and prefer a structured detector when structure is what the sample needs.

\subsection{Selector (Agent 1)}
At each step the Selector is given the common-view evidence for the detectors already run on the current image, the names and one-line mechanism descriptions of the not-yet-run detectors, the remaining budget, the Reporter's suggested next detector and uncertain modality from the previous step, the Analyst's reliability weights, and any retrieved taxonomy entries. The descriptions state only what each detector measures, not which performs better or which domain suits it, so the choice rests on the evidence and the measured reliability rather than on a prior the language model carries from pretraining. The not-yet-run options are presented in a rotated order that advances by one position on every call: a fixed order would invite the model to favor whichever option it sees first, and a random order would break reproducibility, so the rotation is deterministic given the call sequence while giving no detector the first position more often than any other. The Selector returns a JSON object naming the next detector or a stop decision, with a one-sentence justification appended to a shared reasoning trace. A model-written name is matched to the pool up to spacing, underscores, hyphens, and case, and a name outside the pool is treated as a failed call, which triggers the fixed-priority fallback above.

\subsection{Reporter (Agent 2)}
The Reporter receives the common-view evidence and narratives for every detector run so far on the current image and returns two things: a natural-language triage report, and a compact structured manifest holding a verdict (in-distribution, out-of-distribution, or uncertain), a confidence, a dominant failure modality, an uncertain-modality flag, a suggested next detector, and a one-sentence note. Only the manifest is passed back to the Selector. Phi-3.5 is used here rather than the larger Selector model because its tighter output keeps the manifest that the Selector must parse reliably compact. The Reporter is invoked once per detector step and issues no detector calls of its own. If its call fails or returns no manifest, a fallback manifest is built from the accumulated scores so the Selector always receives feedback: the fallback verdict follows the mean calibrated score, its confidence grows with the distance of that mean from the decision point, and its dominant modality is a majority vote over the per-detector dominant modalities.

\subsection{Analyst (Agent 3)}
The Analyst plays two separate roles. Before an image it retrieves relevant prior failures, and this retrieval is CPU-only and calls no language model: it keys on the sample's dominant modality and caption content and returns the most frequently seen matching entries as corrective guidance for the Selector. After a scored image whose ground-truth label is available, and only when the system was wrong, it calls Phi-3.5 to read the reasoning trace of the Selector and Reporter, classify the failure (visual-anchored, text-anchored, cross-modal, or diffuse), and write one sentence of corrective guidance aimed at the Selector. The result is merged into a persistent taxonomy: an entry records the failure type, a description, the visual and caption context, the true and predicted labels, the detectors whose scores were diagnostic, a running count, and a timestamp. A new failure that matches an existing entry by type and description overlap increments that entry and refreshes its guidance rather than adding a duplicate. If the Analyst call fails, the taxonomy is left unchanged for that image, since a missed entry is less costly than a halted run. Separately, the Analyst supplies each detector's reliability weight $w_m=\max\!\bigl(0,\;2\,A_m(\mathcal{C})-1\bigr)$ from its calibration AUROC $A_m(\mathcal{C})$ on the disjoint labeled calibration set, which the aggregation below uses.

\subsection{Pipeline, Aggregation, and Verdict}
For one image the pipeline runs the closed loop: the Analyst retrieves prior failures, the Selector chooses a detector, the detector is scored, the Reporter summarizes the evidence so far and suggests a next detector, and the Selector reads that suggestion and either continues or stops, until the Selector stops or the budget $B$ is spent. Each consulted detector contributes its calibrated position, the fraction of in-distribution calibration scores that fall below its raw score, so every detector is judged against its own in-distribution distribution and all positions share the $[0,1]$ scale on which higher means more out-of-distribution. The system score is the weighted mean of these positions over the detectors the agent consulted, with weights $w_m$: a detector the calibration set found no better than chance receives zero weight and drops out, and a detector it found inverted contributes nothing rather than corrupting the score. When no calibration weights are available, in the configurations without the Analyst, the score is the unweighted mean of the positions. When weights are available but every consulted detector has zero weight, the score is set to the midpoint, since the honest reading of no trusted evidence is no opinion. This weighted average of calibrated positions, and not the order in which detectors were selected, is what produces the recovery reported in the main paper. The agent may stop before the budget is spent once at least two detectors have been consulted and they agree on the calibrated decision with each of them confident, that is, each sitting clear of the decision point by a set margin. In the configurations without the Reporter, the binary verdict applies the calibrated false-positive-rate-$5\%$ threshold: the image is flagged out-of-distribution when the mean position reaches the $0.95$ point.

Beyond this ladder, SABRE's deployment mode adds a fast selective gate, the ``Selective'' in the method's name. Before the agent loop, the best standalone calibrated score across the pool is computed, and an image whose best score falls below a threshold ($\theta=0.15$) is classified in-distribution without invoking any agent, so agent budget is spent only on images near the decision boundary. The gate trades a small amount of accuracy for lower average latency. It is a latency option layered on the full system rather than a rung of the component ladder, and it is therefore not a separate column in the component-analysis tables.

\begin{table}[t]
\centering
\caption{Per-role agent configuration. All roles decode greedily (temperature $0$, top-$k=1$, top-$p=1$, fixed seed). Token caps are sized to each role's output.}
\label{tab:supp-agent-config}
\begin{tabular}{lccc}
\toprule
Role & Model & Max new tokens & Decoding \\
\midrule
Selector & Llama\,3.1\,8B & 96  & greedy \\
Reporter & Phi-3.5       & 256 & greedy \\
Analyst  & Phi-3.5       & 256 & greedy \\
\bottomrule
\end{tabular}
\end{table}

\section{Extended Experimental Results}
\label{sec:supp-experiments}
\subsection{Component Analysis, Full Table}
\label{sec:supp-component}
Table~\ref{tab:supp-agent-histo} gives the per-configuration AUROC underlying the component analysis of the main paper, including both CLIP ViT-B/16 and ViT-L/14 and the GCHTID difficulty ceiling.
\begin{table}[t]
\centering
\caption{Component analysis on histopathology (test split, $B=3$). Each column adds one mechanism. The reliability-calibration step (\textsc{+Mem.}) produces the recovery, and the best and worst single detector bound the range a fixed choice could occupy. $^{\ast}$ marks $p<0.001$ against \textsc{+Rep.}\ GCHTID is a difficulty ceiling.}
\label{tab:supp-agent-histo}
\small
\setlength{\tabcolsep}{4pt}
\begin{tabular}{@{}llccccc@{}}
\toprule
Enc. & Dataset & Sel. & +Rep. & +Mem. & +Crit. & Best/Worst \\
\midrule
\multirow{3}{*}{B/16}
& NCT-CRC & 47.4 & 51.8 & \textbf{76.3}$^{\ast}$ & 76.2 & 79.9/32.9 \\
& CRC-VAL & 49.5 & 52.1 & \textbf{64.6}$^{\ast}$ & 64.4 & 69.6/40.1 \\
& GCHTID  & 43.6 & 45.7 & \textbf{51.4} & \textbf{51.4} & 52.1/40.7 \\
\midrule
\multirow{3}{*}{L/14}
& NCT-CRC & 68.5 & 71.1 & \textbf{84.4}$^{\ast}$ & 84.0 & 85.9/49.6 \\
& CRC-VAL & 73.0 & 75.3 & 81.5 & \textbf{81.9} & 80.4/50.5 \\
& GCHTID  & 55.1 & 55.5 & 56.2 & \textbf{58.0} & 60.7/49.6 \\
\bottomrule
\end{tabular}
\end{table}
\subsection{Detector Benchmark on All Encoders}
\label{sec:supp-detector-all}
\begin{figure*}[t]
\centering
\includegraphics[width=0.49\textwidth]{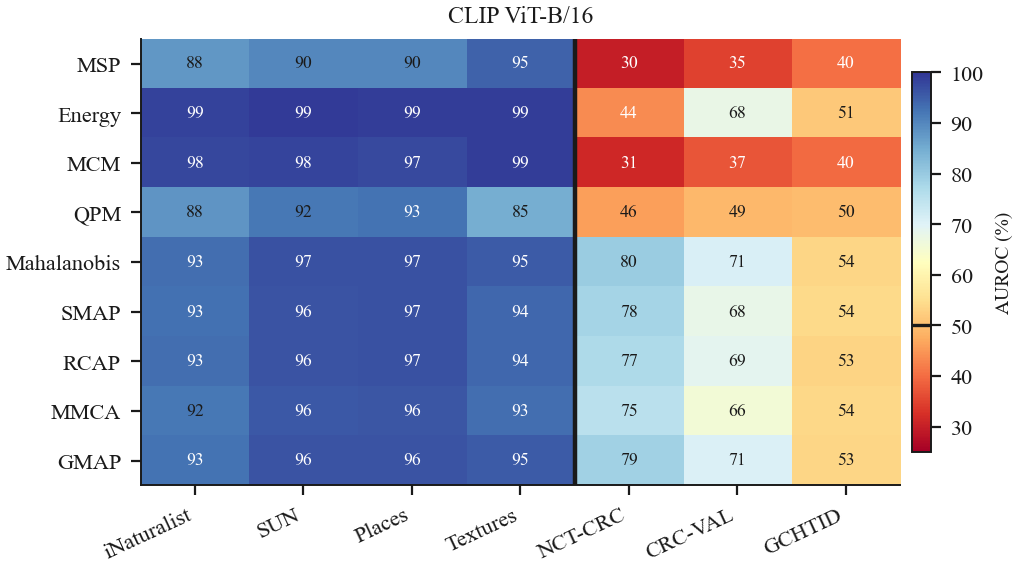}\hfill
\includegraphics[width=0.49\textwidth]{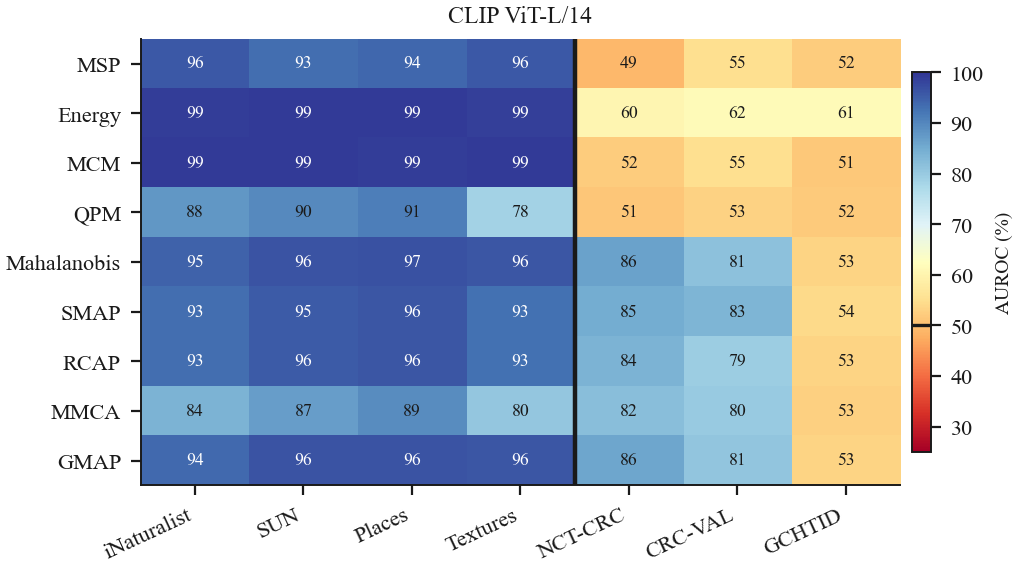}\\[3pt]
\includegraphics[width=0.49\textwidth]{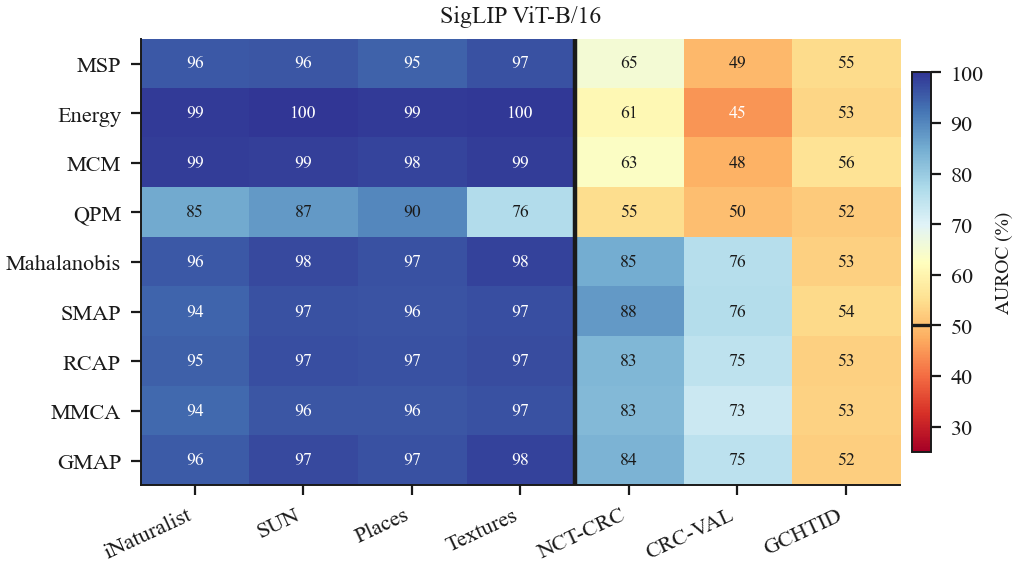}\hfill
\includegraphics[width=0.49\textwidth]{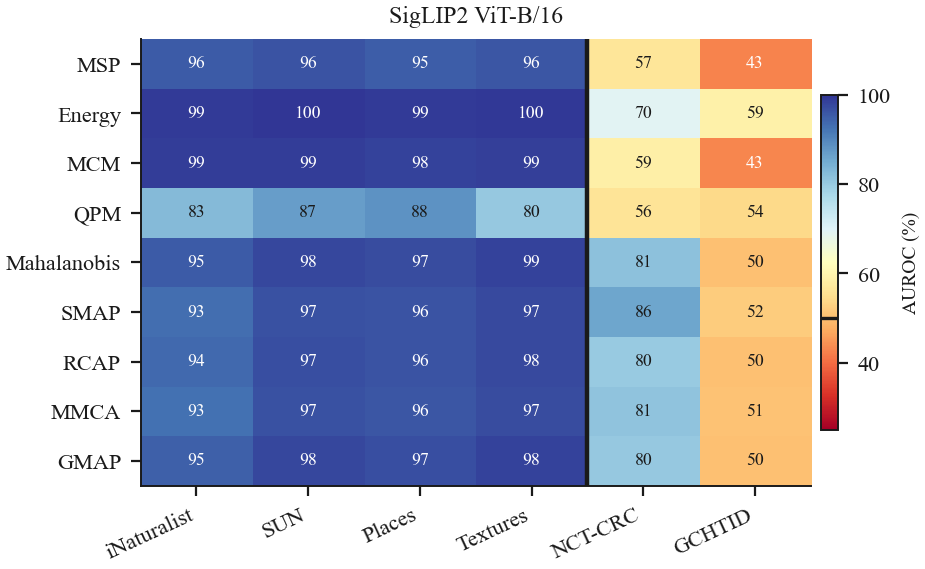}\\[3pt]
\includegraphics[width=0.49\textwidth]{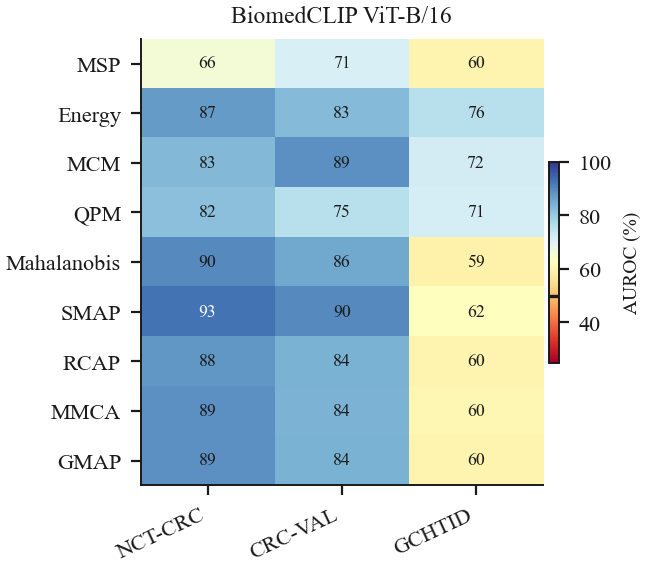}
\caption{Per-detector AUROC (\%) for all five encoders (mean of five seeds), natural
and tissue blocks separated by the rule. On every general-purpose encoder the
confidence and concept detectors invert on tissue while the density detectors
survive, while BiomedCLIP (bottom) lifts the tissue scores but the leading detector still
varies by dataset. SigLIP2 was not run on CRC-VAL at submission.}
\label{fig:supp-heatmaps}
\end{figure*}

Figure~\ref{fig:supp-heatmaps} shows the full per-detector AUROC grid for each of the five encoders, and Tables~\ref{tab:supp-detector-l14}--\ref{tab:supp-detector-biomed} give the underlying numbers (validation split, mean over five seeds). CLIP ViT-B/16 is reported in the main paper.
Two patterns hold across every encoder. The best detector is domain-dependent: a confidence- or concept-matching detector (Energy or MCM) leads on natural images, while a density detector leads on tissue.
The confidence and concept detectors collapse toward or below $50$ on tissue while the density detectors remain informative.
The specialized BiomedCLIP encoder (Table~\ref{tab:supp-detector-biomed}) lifts every detector on tissue and narrows the gap, but does not remove the regime dependence of which detector leads.
\begin{table*}[t]
\centering
\caption{Detector AUROC, CLIP ViT-L/14 (validation, five seeds). Best per row in \textbf{bold}.}
\label{tab:supp-detector-l14}
\small
\begin{tabular}{lccccccccc}
\toprule
& \multicolumn{5}{c}{\textbf{Baselines}} & \multicolumn{4}{c}{\textbf{Ours}}\\
\cmidrule(lr){2-6}\cmidrule(lr){7-10}
Dataset & MSP & Energy & MCM & Maha. & QPM & SMAP & RCAP & MMCA & GMAP \\
\midrule
\multicolumn{10}{l}{\emph{Natural images (ImageNet-100 vs.)}}\\
iNaturalist & 95.8 & 98.9 & \textbf{99.3} & 94.6 & 87.8 & 93.2 & 93.2 & 83.8 & 93.8 \\
SUN & 93.0 & \textbf{99.4} & 99.3 & 96.5 & 89.7 & 95.5 & 95.5 & 87.0 & 96.2 \\
Places & 93.9 & \textbf{99.3} & 98.9 & 96.6 & 91.1 & 96.0 & 96.0 & 89.3 & 96.4 \\
Textures & 95.6 & 98.8 & \textbf{99.3} & 96.1 & 78.4 & 92.8 & 92.9 & 80.4 & 95.9 \\
\midrule
\multicolumn{10}{l}{\emph{Histopathology}}\\
NCT-CRC & 49.3 & 60.4 & 52.1 & \textbf{86.3} & 51.3 & 84.9 & 83.6 & 81.9 & 85.8 \\
CRC-VAL & 55.0 & 61.6 & 55.0 & 81.3 & 53.1 & \textbf{83.4} & 79.3 & 80.0 & 80.9 \\
GCHTID & 52.0 & \textbf{61.4} & 51.4 & 53.3 & 51.7 & 54.1 & 53.4 & 52.7 & 53.3 \\
\bottomrule
\end{tabular}
\end{table*}
\begin{table*}[t]
\centering
\caption{Detector AUROC, SigLIP ViT-B/16 (validation, five seeds). Best per row in \textbf{bold}.}
\label{tab:supp-detector-siglip}
\small
\begin{tabular}{lccccccccc}
\toprule
& \multicolumn{5}{c}{\textbf{Baselines}} & \multicolumn{4}{c}{\textbf{Ours}}\\
\cmidrule(lr){2-6}\cmidrule(lr){7-10}
Dataset & MSP & Energy & MCM & Maha. & QPM & SMAP & RCAP & MMCA & GMAP \\
\midrule
\multicolumn{10}{l}{\emph{Natural images (ImageNet-100 vs.)}}\\
iNaturalist & 95.7 & \textbf{99.3} & 98.9 & 95.8 & 85.3 & 94.3 & 94.8 & 93.7 & 95.6 \\
SUN & 96.0 & \textbf{99.7} & 98.7 & 97.5 & 87.4 & 96.6 & 96.8 & 96.1 & 97.5 \\
Places & 94.6 & \textbf{99.2} & 98.0 & 96.7 & 89.9 & 96.3 & 96.5 & 96.0 & 96.7 \\
Textures & 96.7 & \textbf{99.7} & 99.0 & 98.4 & 76.3 & 97.0 & 97.2 & 96.6 & 98.4 \\
\midrule
\multicolumn{10}{l}{\emph{Histopathology}}\\
NCT-CRC & 65.3 & 60.9 & 63.3 & 84.8 & 54.8 & \textbf{87.6} & 83.1 & 82.7 & 83.8 \\
CRC-VAL & 49.2 & 44.7 & 48.4 & 75.7 & 50.0 & \textbf{76.2} & 74.5 & 72.8 & 75.1 \\
GCHTID & 54.5 & 53.5 & \textbf{56.0} & 52.6 & 51.7 & 54.1 & 52.7 & 53.0 & 52.4 \\
\bottomrule
\end{tabular}
\end{table*}
\begin{table*}[t]
\centering
\caption{Detector AUROC, SigLIP2 ViT-B/16 (validation, five seeds). Best per row in \textbf{bold}. CRC-VAL was not available for this encoder at submission.}
\label{tab:supp-detector-siglip2}
\small
\begin{tabular}{lccccccccc}
\toprule
& \multicolumn{5}{c}{\textbf{Baselines}} & \multicolumn{4}{c}{\textbf{Ours}}\\
\cmidrule(lr){2-6}\cmidrule(lr){7-10}
Dataset & MSP & Energy & MCM & Maha. & QPM & SMAP & RCAP & MMCA & GMAP \\
\midrule
\multicolumn{10}{l}{\emph{Natural images (ImageNet-100 vs.)}}\\
iNaturalist & 95.5 & \textbf{99.3} & 99.1 & 95.4 & 82.6 & 93.1 & 93.6 & 92.8 & 95.0 \\
SUN & 96.5 & \textbf{99.8} & 98.9 & 97.8 & 86.9 & 96.7 & 97.0 & 96.6 & 97.7 \\
Places & 95.3 & \textbf{99.3} & 98.3 & 97.1 & 88.4 & 96.3 & 96.4 & 96.1 & 97.0 \\
Textures & 96.0 & \textbf{99.7} & 98.8 & 98.5 & 80.2 & 97.4 & 97.6 & 97.2 & 98.5 \\
\midrule
\multicolumn{10}{l}{\emph{Histopathology}}\\
NCT-CRC & 56.6 & 69.5 & 59.0 & 81.2 & 56.2 & \textbf{85.7} & 79.9 & 81.0 & 80.3 \\
GCHTID & 42.7 & \textbf{59.2} & 43.0 & 50.4 & 54.2 & 52.1 & 50.5 & 50.5 & 50.4 \\
\bottomrule
\end{tabular}
\end{table*}
\begin{table}[t]
\centering
\caption{Detector AUROC, BiomedCLIP ViT-B/16 (validation, five seeds), histopathology only. Best per row in \textbf{bold}. The domain-specialized encoder lifts every detector but the leading detector still varies by dataset.}
\label{tab:supp-detector-biomed}
\small
\setlength{\tabcolsep}{2.5pt}
\begin{tabular}{@{}lccccccccc@{}}
\toprule
Dataset & MSP & En. & MCM & Ma. & QP. & SM. & RC. & MM. & GM. \\
\midrule
NCT-CRC & 65.7 & 87.4 & 83.0 & 89.6 & 81.7 & \textbf{92.5} & 87.9 & 88.6 & 88.6 \\
CRC-VAL & 70.8 & 82.6 & 88.8 & 85.6 & 75.5 & \textbf{89.6} & 84.1 & 83.7 & 84.2 \\
GCHTID & 60.0 & \textbf{75.6} & 71.9 & 59.2 & 71.4 & 62.3 & 59.9 & 60.5 & 60.1 \\
\bottomrule
\end{tabular}
\end{table}
\subsection{Agent Component Analysis: Significance and Validation Split}
\label{sec:supp-agent-full}
The per-configuration test AUROC underlying the component analysis, for both CLIP ViT-B/16 and ViT-L/14 and including the GCHTID difficulty ceiling, is Table~\ref{tab:supp-agent-histo}.
Table~\ref{tab:supp-agent-paired} reports the paired-bootstrap significance of each configuration step on the test split.
The reliability-weighting step (\textsc{with-memory} over \textsc{with-reporter}) is significant at $p<0.01$ on every histopathology dataset where a detector inverts, and is the largest single step by a wide margin.
The Reporter step and the corrective-guidance step are individually smaller and not always significant, consistent with their role in refining rather than rescuing selection.
Table~\ref{tab:supp-agent-val} gives the validation-split system AUROC, which tracks the test split closely and confirms the configuration ordering was not selected on the test data.
\begin{table}[t]
\centering
\caption{Paired-bootstrap AUROC gains between successive configurations (histopathology, test split, $2000$ resamples). $\Delta$ is the mean paired difference, and the reliability-weighting step dominates. Natural-image steps are within noise of zero and omitted.}
\label{tab:supp-agent-paired}
\small
\begin{tabular}{llccc}
\toprule
Enc. & Dataset & Step & $\Delta$ & $p$ \\
\midrule
\multirow{2}{*}{B/16}
& NCT-CRC & +Memory & $+24.5$ & $<0.001$ \\
& CRC-VAL & +Memory & $+12.5$ & $<0.001$ \\
\midrule
\multirow{2}{*}{L/14}
& NCT-CRC & +Memory & $+13.2$ & $<0.001$ \\
& CRC-VAL & +Memory & $+6.2$ & $0.002$ \\
\midrule
Biomed & NCT-CRC & +Memory & $+8.5$ & $<0.001$ \\
\bottomrule
\end{tabular}
\end{table}
\begin{table}[t]
\centering
\caption{Agent system AUROC on the validation split (histopathology, CLIP ViT-B/16, budget $B=3$), with the best single detector for reference. The ordering and magnitude of the configuration steps match the test split (Table~\ref{tab:supp-agent-histo}).}
\label{tab:supp-agent-val}
\small
\begin{tabular}{lccccc}
\toprule
Dataset & Sel. & +Rep. & +Mem. & +Crit. & Best \\
\midrule
NCT-CRC & 48.8 & 51.7 & 74.4 & \textbf{75.1} & 80.1 \\
CRC-VAL & 46.7 & 49.2 & 64.4 & \textbf{65.5} & 71.3 \\
GCHTID  & 44.8 & 45.5 & 49.5 & \textbf{52.2} & 53.3 \\
\bottomrule
\end{tabular}
\end{table}
\subsection{Detector Selection Counts}
\label{sec:supp-selection}
Table~\ref{tab:supp-selection-tissue} gives the detector selection share across configurations on tissue, and Table~\ref{tab:supp-selection-natural} on natural images. As calibration engages on tissue, selection moves toward the density detectors that are sound there (Mahalanobis $0.5\%\!\to\!16.4\%$, SMAP $0\%\!\to\!7.6\%$) and away from the inverted ones (QPM $8.7\%\!\to\!0$, MCM halved). Detectors that keep a high selection share but a zero weight (MSP, MCM) are consulted but carry no influence in the aggregate, so effective reliance is read from the weight, not the share. On natural images the distribution barely moves, because no detector needs ruling out: the confidence and concept detectors that the tissue regime suppresses are exactly the ones that are reliable here.
\begin{table}[t]
\centering
\caption{Detector selection share (\%) across configurations on tissue (NCT-CRC, CLIP ViT-B/16 test), with calibration weight $w_m$.}
\label{tab:supp-selection-tissue}
\small
\begin{tabular}{lccccc}
\toprule
& \multicolumn{4}{c}{Selection share (\%)} & \\
\cmidrule(lr){2-5}
Detector & Sel. & +Rep. & +Mem. & +Crit. & $w_m$ \\
\midrule
\multicolumn{6}{l}{\emph{Inverted on tissue}}\\
MSP    & 30.6 & 32.6 & 32.3 & 31.5 & 0.00 \\
MCM    & 27.8 & 23.8 & 10.6 & 11.4 & 0.00 \\
QPM  & 8.7  & 0.2  & 0.0  & 0.0  & 0.00 \\
Energy & 0.1  & 0.0  & 0.0  & 0.0  & 0.00 \\
\midrule
\multicolumn{6}{l}{\emph{Sound on tissue}}\\
Mahalanobis & 0.5  & 2.0  & 16.4 & 16.5 & 0.48 \\
SMAP        & 0.0  & 4.5  & 7.6  & 6.8  & 0.48 \\
MMCA        & 28.1 & 29.8 & 31.1 & 31.8 & 0.44 \\
GMAP        & 4.3  & 7.1  & 2.1  & 2.0  & 0.45 \\
RCAP        & 0.0  & 0.0  & 0.0  & 0.0  & 0.46 \\
\bottomrule
\end{tabular}
\end{table}
\begin{table}[t]
\centering
\caption{Detector selection share (\%) across configurations on a natural set (ImageNet-100 vs.\ iNaturalist, CLIP ViT-B/16 test), with calibration weight $w_m$. Unlike the tissue regime, the distribution is stable and the confidence and concept detectors retain weight, because none is inverted here.}
\label{tab:supp-selection-natural}
\small
\begin{tabular}{lccccc}
\toprule
& \multicolumn{4}{c}{Selection share (\%)} & \\
\cmidrule(lr){2-5}
Detector & Sel. & +Rep. & +Mem. & +Crit. & $w_m$ \\
\midrule
MSP         & 31.0 & 34.2 & 36.7 & 36.3 & 0.70 \\
MCM         & 25.0 & 19.5 & 26.0 & 26.7 & 0.98 \\
Energy      & 0.0  & 0.0  & 2.2  & 2.5  & 0.99 \\
Mahalanobis & 1.7  & 2.4  & 0.0  & 0.1  & 0.89 \\
SMAP        & 0.0  & 3.7  & 12.4 & 11.9 & 0.90 \\
MMCA        & 29.3 & 32.7 & 19.5 & 19.5 & 0.89 \\
GMAP        & 5.0  & 7.3  & 3.0  & 2.8  & 0.88 \\
QPM       & 8.0  & 0.2  & 0.2  & 0.2  & 0.90 \\
\bottomrule
\end{tabular}
\end{table}

\subsection{Budget Usage}
\label{sec:supp-budget}
Table~\ref{tab:supp-budget} reports the distribution of detector calls per image.

\begin{center}
\captionof{table}{Budget usage by configuration (NCT-CRC, CLIP ViT-B/16 test, $B=3$). Mean calls per image, the fraction of images that reach the budget, and the two- and three-call counts.}
\label{tab:supp-budget}
{\small
\begin{tabular}{lcccc}
\toprule
Configuration & Mean & Saturated & 2 calls & 3 calls \\
\midrule
selector-only & 2.56 & 56\% & 222 & 278 \\
with-reporter & 2.59 & 59\% & 203 & 297 \\
with-memory   & 2.62 & 62\% & 192 & 308 \\
with-critique & 2.62 & 62\% & 192 & 308 \\
\bottomrule
\end{tabular}}
\end{center}

Under a budget of $B=3$, the mean call count is between $2.4$ and $2.7$ across configurations, and the system stops before saturating the budget on roughly $40\%$ of images, so it is selecting within the budget rather than merely exhausting it.

\end{document}